\documentclass[a4paper,oneside,final,notitlepage,onecolumn,12pt]{article}
\usepackage{amsfonts}
\usepackage{epsf}
\usepackage{graphicx}
\usepackage{amssymb,eso-pic}
\usepackage{latexsym}
\usepackage{tabularx}
\usepackage{amsxtra} 
\usepackage{hyperref}
\usepackage{t1enc}
\usepackage{amsmath,accents}
\usepackage{ifpdf,epsfig,array,amsmath,amssymb}

\usepackage{psfrag} 
\psfrag{na}{\footnotesize $n^a$}   
\psfrag{ta}{\footnotesize $\sigma^a$}   
\psfrag{vna}{\footnotesize $\check{n}^a$}
\psfrag{vNa}{\footnotesize $\check{N}^a$}
\psfrag{scrW}{\footnotesize $\mycal{W}_{\rho_0}$}
\psfrag{t=t1}{\footnotesize $\Sigma_{\sigma_1}$}
\psfrag{t=t2}{\footnotesize $\Sigma_{\sigma_2}$}
\psfrag{r=all}{\footnotesize $\rho=const$}
\psfrag{hna}{\footnotesize $\hat{n}^a$}
\psfrag{tna}{\footnotesize $\tilde{n}^a$}
\psfrag{hab,Kab}{\footnotesize $h_{ab}, K_{ab}$}
\psfrag{thab,tKab}{\footnotesize $\tilde h_{ab}, \tilde K_{ab}$}
\psfrag{hhab,hKab}{\footnotesize $\hat h_{ab}, \hat K_{ab}$}
\psfrag{chab,cKab}{\footnotesize $\check{h}_{ab}, \check{K}_{ab}$}

\usepackage[normalem]{ulem}
\usepackage{color}
\definecolor{blue}{rgb}{0,0,1}
\definecolor{red}{rgb}{1,0,0}

\newcommand{\interior}[1]{\accentset{\smash{\raisebox{-0.12ex}{$\scriptstyle\circ$}}}{#1}\rule{0pt}{2.3ex}}
\makeatother

\DeclareFontFamily{OT1}{rsfs}{} \DeclareFontShape{OT1}{rsfs}{m}{n}{
<-7> rsfs5 <7-10> rsfs7 <10-> rsfs10}{}
\DeclareMathAlphabet{\mycal}{OT1}{rsfs}{m}{n}

\def\sc{{\hskip 3.5pt {{}^{{}^{{}_{{}_{\bowtie}}}}} \kern -8.pt{}}}  
\def\SC{{\hskip 3.5pt {{}^{{}^{{}^{{}_{{}_{\bowtie}}}}}} \kern -10.5pt{}}}

\DeclareMathAlphabet{\mathpzc}{OT1}{pzc}{m}{it}

\begin{document}

\newtheorem{theorem}{Theorem}[section]
\newtheorem{lemma}{Lemma}[section]
\newtheorem{proposition}{Proposition}[section]
\newtheorem{corollary}{Corollary}[section]
\newtheorem{conjecture}{Conjecture}[section]
\newtheorem{condition}{Condition}[section]
\newtheorem{example}{Example}[section]
\newtheorem{definition}{Definition}[section]
\newtheorem{remark}{Remark}[section]
\newtheorem{exercise}{Exercise}[section]
\newtheorem{axiom}{Axiom}[section]
\renewcommand{\theequation}{\thesection.\arabic{equation}} 

\author{{Istv\'an R\'acz}\,\thanks{ ~email: racz.istvan@wigner.mta.hu}  
\\ \\ 
{Wigner RCP}\\  {H-1121 Budapest, Konkoly Thege Mikl\'os \'ut 29-33. Hungary
}}
\title{Dynamical determination of the gravitational degrees of freedom} 

\maketitle

\begin{abstract}
$[n+1]$-dimensional ($n\geq 3$) smooth Einsteinian spaces of Euclidean and Lorentzian signature are considered. The base manifold $M$ is supposed to be smoothly foliated by a two-parameter family of codimension-two-surfaces which are orientable and compact without boundary in $M$. By applying a pair of nested $1+n$ and $1+[n-1]$ decompositions, the canonical form of the metric and the conformal structure of the foliating codimension-two-surfaces a gauge fixing, analogous to the one applied in \cite{racz_geom_cauchy}, is introduced. In verifying that the true degrees of freedom of gravity may conveniently be represented by the conformal structure it is shown first that regardless whether the primary space is Riemannian or Lorentzian, in terms of the chosen geometrically distinguished new variables, the $1+n$ momentum constraint can be written as a first order symmetric hyperbolic system.
It is also argued that in the generic case the Hamiltonian constraint can be solved as an algebraic equation. By combining the $1+n$ constraints with the part of the reduced system of the secondary $1+[n-1]$ decomposition that governs the evolution of the conformal structure---in the Riemannian case with an additional introduction of an imaginary `time'---a well-posed mixed hyperbolic-algebraic-hyperbolic system is formed. It is shown that regardless whether the primary space is of Riemannian or Lorentzian if a regular origin exists solutions to this system are also solutions to the full set of Einstein's equations. This, in particular, offers the possibility of developing a new method for solving Einstein's equations in the Riemannian case.   
The true degrees of freedom of gravity are also found to be subject of a nonlinear wave equation.
\end{abstract} 

\date


\section{Introduction}\label{introduction}
\setcounter{equation}{0}

The determination of the true degrees of freedom is of crucial importance in the proper description of classical and quantum aspects of various dynamical systems in physics. This paper is to provide a dynamical determination of the gravitational degrees of freedom {applicable} in a wide class of metric theories of gravity. 

\medskip
It came as a great surprise, after working out the details of this analysis for the Lorentzian case, that a slightly different but completely analogous argument applies for Einsteinian spaces of Euclidean signature{, and it is also straightforward to extend the argument to spaces of arbitrary dimension}. As a result in this paper $[n+1]$-dimensional ($n\geq 3$) smooth Einsteinian spaces of Euclidean and Lorentzian signature could be considered. The base manifold $M$ is assumed to be smoothly foliated by a two-parameter family of codimension-two-surfaces. These surfaces are {supposed} to be orientable and compact with no boundary in $M$. This topological assumption allows us to apply a pair of nested $1+n$ and $1+[n-1]$ decompositions. {B}ased on the use the canonical form of the metric and the notion of the conformal structure {a gauge choice---analogous to the one made in \cite{racz_geom_cauchy}---}is applied. By restricting attention to generic spaces satisfying Conjecture \ref{conj} { and possessing} a regular origin the following points emerged:

\begin{itemize}
\item For a suitable choice of the dependent variables the $1+n$ constraints can be put into the form an algebraic-hyperbolic system. In particular, 
\begin{itemize}
\item the $1+n$ momentum constraint can be written as a first order symmetric hyperbolic system
\item for generic spaces the Hamiltonian constraint can be solved as an algebraic equation. 
\end{itemize}
\item By combining the $1+n$ constraints with the trace free projection of the reduced evolutionary system of the secondary $1+[n-1]$ decomposition---in the Riemannian case with an additional introduction of an imaginary `time'---a mixed hyperbolic-algebraic-hyperbolic system is formed.
\item It is shown---provided that a regular origin exists---that solutions to this mixed hyperbolic-algebraic-hyperbolic system are also solutions to the full set of Einstein's equations.
\item Remarkably the analysis suggests an efficient new method to get solutions of Einstein's equation in the Riemannian case.
\item The true degrees of freedom of {the involved} metric theories of gravity are found to be {conveniently} represented by the conformal structure of the foliating codimension-two-surfaces which are subject of a nonlinear wave equation.
\end{itemize}

A key observation one has to make before one can come to grips with the above statements is that the constraint equations are always underdetermined, so they may change their character if a suitable new set of variables {is chosen}. As a trivial example one may think of the equation 
\begin{equation}
\partial_x^2 f + \partial_y^2 f + g = 0
\end{equation}
as an elliptic or algebraic equation depending on whether it is to be solved for $f$ or $g$, respectively. 

\medskip

An unexpected extra bonus originates from the use of the proposed {geometrically distinguished variables}. Namely, most of the new results apply on equal footing---if not then with relatively slight technical differences---to the Riemannian and Lorentzian cases. The principal parts of the equations either remain intact, as in case of the constraint equations, or can be put into a convenient common form independent of the signature of the base manifold.

\medskip

Th{is} paper is {structured} as follows. In Section \ref{prelim} the kinematic setup is fixed by spelling out our geometrical and topological assumptions.  Section \ref{1+(1+[n-1])_decomp} is to  introduce the nested $1+n$ and $1+[n-1]$ decompositions. The applied gauge fixing, along with the inclusion of the most important relations in terms of the chosen geometrically distinguished variables, is presented in Section \ref{spec_choice_n}. Section \ref{constraints} is to show that the $1+n$ constraints form a well-posed algebraic-hyperbolic system. The fundamental relations between various projections of Einstein's equation are explored in Section \ref{evolutionary}. The solubility of the reduced mixed system is discussed in Section \ref{reduced}.  {Finally,} in Section \ref{final} {some of the implications of the proposed new approach are discussed}. 

\section{Preliminaries}\label{prelim}
\setcounter{equation}{0}

As in \cite{racz_geom_det} {Einsteinian spaces represented by} a pair $(M,g_{ab})$ {will be considered}, where $M$ is an $[n+1]$-dimensional ($n\geq 3$) smooth, paracompact, connected, orientable manifold endowed with a smooth metric $g_{ab}$ with Euclidean or Lorentzian signature\,\footnote{All of our other conventions will be as in \cite{wald}.} subject to Einstein's  equations
\begin{equation}\label{geom_gd}
G_{ab}-\mycal{G}_{ab}=0\,,
\end{equation}
where the source term $\mycal{G}_{ab}$ is assumed to have vanishing divergence. 

\medskip

{T}he {latter} assumption {holds for} Einstein-matter systems with
\begin{equation}
\mycal{G}_{ab} = 8\pi\,T_{ab} - \Lambda\,g_{ab}\,,
\end{equation}
where $T_{ab}$ {stands for} the energy-momentum tensor{ of matter fields satisfying their field equations, and} $\Lambda$ {is} the cosmological constant. {For further discussion on the implications of (\ref{geom_gd}) see Refs.}~\cite{racz_geom_det, racz_geom_cauchy}.

\medskip

In arranging the arena for the nested $1+n$ and $1+[n-1]$ decompositions we shall assume first that the manifold $M$ is foliated by a one-parameter family of hypersurfaces, i.e.~$M\simeq\mathbb{R}\times\Sigma$, for some codimension-one manifold $\Sigma$. 
Note that this assumption hold{s} for globally hyperbolic spacetimes \cite{geroch}. {Nevertheless,} as the signature of the metric need not to be Lorentzian or even if it was, {it} may not {be necessary} to assume global hyperbolicity of the pertinent spacetime. Note also that our assumption is equivalent to the existence of a smooth function $\sigma:M\rightarrow \mathbb{R}$ with non-vanishing gradient $\nabla_a \sigma$ such that the $\sigma=const$ level surfaces $\Sigma_{\sigma}=\{\sigma\}\times \Sigma$ comprise the one-parameter foliation of $M$.  {As noted already the space $(M,g_{ab})$ may not have anything to do with time evolution. Nevertheless, a vector field $\sigma^a$ on $M$ will be referred as `evolution vector field' if the relation $\sigma^e\nabla_e\sigma=1$ holds. Notice that this condition guaranties that $\sigma^a$ nowhere vanishes nor becomes 
tangent to the $\sigma=const$ level surfaces.}

\medskip

To have the desired two-parameter family of homologous codimension-two-surfaces it suffices to guarantee that on one of the $\sigma=const$ level surfaces---say on $\Sigma_{0}$---there exists a smooth function $\rho: \Sigma_{0}\rightarrow \mathbb{R}$, with nowhere vanishing gradient, such that the $\rho=const$ level surfaces $\mycal{S}_\rho$ are homologous {codimension-two-surfaces}, and they are orientable compact {and} without boundary in $M$.

\medskip

{We get then the desired two-parameter family of homologous codimension-two-surfaces $\mycal{S}_{\sigma,\rho}$ by the Lie transport of the chosen foliation of $\Sigma_{0}$ along the integral curves of the vector field  $\sigma^a$.}

\medskip

{By a straightforward adaptation of the procedure described in details in the last two paragraphs of section 2 of \cite{racz_geom_cauchy} one may introduce smooth (local) coordinates $(\sigma,\rho,x^3,\dots,x^{n+1})$ adopted to both of the vector fields $\sigma^a$ and $\rho^a$, where the latter is supposed to be everywhere tangential to the $\sigma=const$ level surfaces, and chosen such that  $\rho^e\nabla_e\rho=1$ throughout $M$.}

\section{The nested $1+n$ and $1+[n-1]$ decompositions}\label{1+(1+[n-1])_decomp} 
\setcounter{equation}{0}

In performing first the $1+n$ splitting denote by $n^a$ the `unit norm' vector field that is normal to the $\sigma=const$ level surfaces. To adjust our setup in order to be able to {accommodate both Riemannian} and Lorentzian {spaces} simultaneously the sign of the norm of $n^a$ will not be fixed. {I}t will be assumed that
\begin{equation}
n^a n_a=\epsilon\,,
\end{equation}
where $\epsilon$ takes the value $+1$ and $-1$, respectively. 

\medskip

As in \cite{racz_geom_det} the induced metric $h_{ab}$ and the pertinent projection operator ${h^a}_b$ on the level surfaces of $\sigma:M\rightarrow \mathbb{R}$ are then given as
\begin{equation}
h_{ab}=g_{ab}-\epsilon\, n_a n_b \hskip3mm {\rm and} \hskip3mm 
{h^a}_b={g^a}_b -\epsilon\,  n^a n_b\,,
\end{equation}
respectively.

\medskip

The unit normal $n^a$ to the {$\sigma=const$} level surfaces may {then} be decomposed as 
\begin{equation}
n^a=\frac1N\,\left[ (\partial_\sigma)^a-N^a\right]\,,
\end{equation}
where the `laps' and `shift'{, $N$ and $N^a$,} of the `evolution' vector field $\sigma^a=(\partial_\sigma)^a$ {are given} as 
\begin{equation}
N= \epsilon\,(\sigma^e n_e) \hskip0.5cm {\rm and} \hskip0.5cm N^a={h^{a}}_{e}\,\sigma^e\,,
\end{equation}
respectively.

\medskip

{T}he Hamiltonian and momentum constraints and the evolution equation can {then} be given---see (3.7), (3.8) and (3.12) of \cite{racz_geom_det}---as 
\begin{align} 
E^{{}^{(\mathcal{H})}} = {}& n^e n^f E_{ef}=\tfrac12\,\{ -\epsilon\,{}^{{}^{(n)}}\hskip-1mm R + \left({K^{e}}_{e}\right)^2 - K_{ef} K^{ef} - 2\,\mathfrak{e} \}=0\,, \label{expl_eh}\\
E^{{}^{(\mathcal{M})}}_a  = {}& {h^e}_a n^f  E_{ef}=D_e {K^{e}}_{a} - D_a {K^{e}}_{e} - \epsilon\,\mathfrak{p}_{a}=0\,,\label{expl_em} \\
E^{{}^{(\mathcal{EVOL})}}_{ab}\hskip-.1cm{} = {}& {}^{{}^{(n)}}\hskip-1mm R_{ab} +\epsilon\,\left\{ -\mycal{L}_n K_{ab} - ({K^{e}}_{e}) K_{ab} + 2\,K_{ae}{K^{e}}_{b} - {\epsilon}{N^{-1}}\,D_a D_b N \right\} \nonumber \\
& +\tfrac{1+\epsilon}{(n-1)}\,{h}_{ab}\,E^{{}^{(\mathcal{H})}} -\left(\mathfrak{S}_{ab}-\tfrac1{n-1}\,h_{ab}\,[\mathfrak{S}_{ef}\,h^{ef}+\epsilon\,\mathfrak{e} ] \right)=0\,, \label{evol_ev_2}
\end{align}
where $D_a$ denotes the covariant derivative operator associated with $h_{ab}$, {$K_{ab}$ stands for the extrinsic curvature}
\begin{equation}\label{extcurv}
{K_{ab}= {h^{e}}_{a} \nabla_e n_b=\tfrac12\,\mycal{L}_n  h_{ab}\,, }
\end{equation}
{where $\mycal{L}_n$ denotes the Lie derivative with respect to $n^a$,}
whereas $\mathfrak{e}$, $\mathfrak{p}_{a}$ and $\mathfrak{S}_{ab}$ {stand for the} projections of the source term of (\ref{geom_gd}) {given} as 
\begin{equation}\label{1+n_source}
\mathfrak{e}= n^e n^f\,\mycal{G}_{ef}\,,\hskip4mm \mathfrak{p}_{a}=\epsilon\,{h^{e}}_{a} n^f\, \mycal{G}_{ef}\hskip4mm {\rm and} \hskip4mm \mathfrak{S}_{ab}={h^{e}}_{a}{h^{f}}_{b}\,\mycal{G}_{ef}\,.
\end{equation}

\medskip

Note that (\ref{evol_ev_2}) can also be written as 
\begin{equation}\label{evol_ch0_n}
{}^{{}^{(n)}}\hskip-1mm {G}_{ij}- {}^{{}^{(n)}}\hskip-1mm \mycal{G}_{ij}=0\,,
\end{equation}
where
\begin{eqnarray}\label{int_evol_1+n} 
&& {}^{{}^{(n)}}\hskip-1mm \mycal{G}_{ab} = \mathfrak{S}_{ab} -\epsilon\left\{ -\mycal{L}_n K_{ab} - ({K^{e}}_{e}) K_{ab} + 2\,K_{ae}{K^{e}}_{b} - {\epsilon}\,{N^{-1}}\,D_a D_b N  \right. \\ && \left. \phantom{{}^{{}^{(n)}}\hskip-1mm \mycal{G}_{ab} = \mathfrak{S}_{ab}} + h_{ab}\left[\mycal{L}_n ({K^e}_{e}) + \tfrac12\,({K^{e}}_{e})^2   + \tfrac12\,K_{ef}{K^{ef}} + {\epsilon}\,{N^{-1}}\,D^e D_e N \right]\right\}  \nonumber \,. 
\end{eqnarray}
 
{By making use of the positive definite metric $\hat \gamma_{ij}$, induced on the $\mycal{S}_{\sigma,\rho}$ level surfaces, and the unit norm field} 
\begin{equation}\label{nhat}
{\hat n^i={\hat{N}}^{-1}\,[\, (\partial_\rho)^i-{\hat N}{}^i\,]}
\end{equation}
{normal to the $\mycal{S}_{\sigma,\rho}$ level surfaces within $\Sigma_\sigma$, where $\hat N$ and $\hat N^i$ denotes the `laps' and `shift' of the `evolution' vector field $\rho^i=(\partial_\rho)^i$ on $\Sigma_\sigma$,}
the metric $h_{ij}$ and 
\begin{equation}\label{evol_ch}
{}^{{}^{(n)}}\hskip-1mm E_{ij} = {}^{{}^{(n)}}\hskip-1mm {G}_{ij}- {}^{{}^{(n)}}\hskip-1mm \mycal{G}_{ij}
\end{equation}
can be decomposed as 
\begin{equation}\label{hij}
h_{ij}=\hat \gamma_{ij}+\hat  n_i \hat n_j\,,
\end{equation}
and 
\begin{eqnarray} 
\hat E^{{}^{(\mathcal{H})}} {}&\hskip-.4cm=&\hskip-.2cm{} \tfrac12\,\{ -\hat R + ({{\hat K}{}^{l}}{}_{l})^2 - \hat K_{kl}\hat K^{kl} - 2\,\hat{\mathfrak{e}} \}\,,\label{ham_1+(1+2_n)} \\
\hat E^{{}^{(\mathcal{M})}}_i {}&\hskip-.4cm=&\hskip-.2cm{} \hat D^l {{\hat K}}{}_{li} - \hat D_i {{\hat K}{}^{l}}{}_{l} - \hat{\mathfrak{p}}_{i}\,,\label{mom_1+(1+2_n)}\\
\hat E^{{}^{(\mathcal{EVOL})}}_{ij} {}&\hskip-.4cm=&\hskip-.2cm{} \hat R_{ij} -\mycal{L}_{\hat n} \hat K_{ij} - ({{\hat K}{}^{l}}{}_{l}) {\hat K}_{ij} + 2\,\hat K_{il}{{\hat K}{}^{l}}{}_{j} - {\hat{N}}^{-1}\,\hat D_i \hat D_j \hat N   \nonumber \\ 
&& \phantom{\hskip-.2cm{} \hat R_{ij}}
+ \hat\gamma_{ij}\{\mycal{L}_{\hat n} {\hat K}{}^l{}_{l} + \hat K_{kl}{\hat K}{}^{kl} + {\hat{N}}^{-1}\,\hat D^l \hat D_l \hat N  \} \label{evol_1+(1+2_n)} - [\hat{\mathfrak{S}}_{ij}-\hat{\mathfrak{e}}\,\hat \gamma_{ij}], 
\end{eqnarray}
where $\hat D_i$, $\hat R_{ij}$ and $\hat R$ denote the covariant derivative operator, the Ricci tensor and scalar curvature associated with $\hat \gamma_{ij}$, respectively.\,\footnote{Note that $\Sigma_\sigma$ is spacelike therefore in the secondary $1+[n-1]$ decomposition the correspondent of $\epsilon$ takes the value $+1$ regardless whether the primary space is of Riemannian or Lorentzian.}
The `hatted' source terms $\hat{\mathfrak{e}}$, $\hat{\mathfrak{p}}_{i}$ and $\hat{\mathfrak{S}}_{ij}$ and the extrinsic curvature $\hat K_{ij}$ are defined as 
\begin{equation}\label{1+n_sourcehat}
\hat{\mathfrak{e}}=\hat n^k \hat n^l\,{}^{{}^{(n)}}\hskip-1mm \mycal{G}_{kl}\,, \hskip4mm \hat{\mathfrak{p}}_{i}={{\hat \gamma}^{k}}{}_{i}\,\hat n^l\, {}^{{}^{(n)}}\hskip-1mm \mycal{G}_{kl} \hskip4mm {\rm and} \hskip4mm \hat{\mathfrak{S}}_{ij}={{\hat \gamma}^{k}}{}_{i} {{\hat \gamma}^{l}}{}_{j}\,{}^{{}^{(n)}}\hskip-1mm \mycal{G}_{kl}\,,
\end{equation}
and
\begin{equation}\label{hatextcurv_n}
\hat K_{ij}= {{\hat \gamma}^{l}}{}_{i}\, D_l\,\hat n_j=\tfrac12\,\mycal{L}_{\hat n} {\hat \gamma}_{ij}\,.
\end{equation}

\section{The gauge choice {and the conformal structure}}\label{spec_choice_n}
\setcounter{equation}{0}

So far {the} $1+n$ and $1+[n-1]$ decompositions were kept to be generic. 

\medskip

In order to reduce the complexity of the field equations we shall assume that the {shift vector $N^a$ of the vector field $\sigma^a=(\partial_\sigma)^a$ is identically zero on $M$.}

\medskip

For a detailed verification of this assumption in the Lorentzian case see subsection 4.1 of \cite{racz_geom_cauchy}. Note, however, that in case of Riemannian spaces one cannot refer to the correspondent of the result of M\"uller and S\'anches \cite{MOSM}. Nevertheless, based on the diffeomorphism invariance of the underlying theory we may simply require, without loss of generality, that the metric $g_{ab}$ possesses the canonical form
\begin{equation}\label{pref_n}
g_{ab}=\epsilon\,N^2\, (d\sigma)_a(d\sigma)_b+h_{ab}\,,
\end{equation} 
with a bounded lapse function $N:M\rightarrow \mathbb{R}$, and with a smooth Riemannian metric $h_{ab}$ on the $\Sigma_{\sigma}$ level surfaces.\,\footnote{Note that in the Riemannian case the canonical form of the metric suggests the use of the imaginary `time' $\varsigma=i\,\sigma$ instead of $\sigma$ as by such a replacement one could formally put the metric of Riemannian spaces into Lorentzian form. Nevertheless, as such a replacement would require simultaneous adjustments of all the $\sigma$-derivatives it appears to be preferable to postpone this to be done in Section \ref{reduced}.\label{foot} There, by applying this replacement, a new method solving Einstein's equations in the Riemannian case will be proposed.} 

\medskip

{Beside this gauge choice we shall assume that there exists a smooth non-vanishing function $\nu:M\rightarrow\mathbb{R}$ relating the two lapse functions as}
\begin{equation}\label{taurhoSection1_n}
{N=\nu\,\hat N\,.}
\end{equation}

\medskip

{We shall also apply a split of the metric $\hat\gamma_{ij}$ into a conformal factor and the conformal structure as }
\begin{equation}\label{thetaphiSection_n}
{\hat\gamma_{ij} = \Omega^2\,\gamma_{ij}\,,}
\end{equation}
{where $\Omega:M\rightarrow\mathbb{R}$ is function---which does not vanish except at an origin where the foliation $\mycal{S}_{\sigma,\rho}$ smoothly reduces to a point on the $\Sigma_\sigma$ level surfaces---whereas the conformal structure $\gamma_{ij}$ is defined such that}
\begin{equation}\label{eta} 
{\gamma^{ij}(\mycal{L}_\eta\gamma_{ij})=0}
\end{equation}
{holds on each of the $\mycal{S}_{\sigma,\rho}$ surfaces, where $\eta^a$ stands for either of the coordinate basis fields  $\sigma^a=(\partial_\sigma)^a$ or $\rho^a=(\partial_\rho)^a$.}

\medskip

The argument applied in subsection 4.1 of \cite{racz_geom_cauchy} {verifying} the {viability of the} geometrically preferred splitting  (\ref{thetaphiSection_n}) of $\hat\gamma_{ij}$ can be used on equal footing to both Riemannian and Lorentzian spaces. The only distinction is that instead of the factors $2$ and $\tfrac12$ in equations (4.6) and (4.8) of \cite{racz_geom_cauchy} the factors $n-1$ and $\tfrac{1}{n-1}$ have to be used, respectively. In particular, (4.6) of \cite{racz_geom_cauchy} reads as 
\begin{equation}\label{eta_2_n}
\hat\gamma^{ij}(\mycal{L}_\eta\hat\gamma_{ij})=\gamma^{ij}(\mycal{L}_\eta\gamma_{ij})+(n-1)\,\mycal{L}_\eta(\ln \Omega^2)\,.
\end{equation} 

\medskip

As discussed in \cite{racz_geom_cauchy} an origin {exists on a $\Sigma_\sigma$ level surface if the foliating codimension-two-surfaces smoothly reduce to a point at the location $\rho=\rho_*$. An origin in $M$ is represented by a line {$\mycal{W}_{\rho_*}$} yielded by the Lie transport of an origin on one of the $\Sigma_\sigma$ level surfaces along one of the evolution vector fields.}

\medskip

An origin {$\mycal{W}_{\rho_*}$} will be referred as regular if {the occurrence of various defects such as the existence of a conical singularity is excluded. If {$\mycal{W}_{\rho_*}$} is regular then} there exist smooth functions $\hat N{}_{(2)}, \Omega{}_{(3)}$ and $\hat N^A_{(1)}$ such that, in a neighborhood of the location $\rho=\rho_*$ on the $\Sigma_\sigma$ level surfaces, the basic variables $\hat N, \Omega$ and $\hat N^A$ can be given as 
\begin{align}\label{origin_n} 
\hat N=1+  (\rho-\rho_*)^2\,\hat N{}_{(2)}, \ \Omega=(\rho-\rho_*)+ (\rho-\rho_*)^3\,\Omega{}_{(3)} \ \ {\rm and} \ \ \hat N^A= (\rho-\rho_*)\,\hat N^A_{(1)}\,.
\end{align}
{Accordingly, $\Omega$ vanishes at a regular origin and the limiting behavior $\hat\gamma^{ij}(\mycal{L}_\rho\hat\gamma_{ij})\rightarrow \pm\infty$, while $\rho\rightarrow \pm\rho_*$, does also signify it.}  

\medskip

In what follows ${}^{{}^{(\gamma)}}\hskip-1mm R$ and $\mathbb{D}_i$ denote the scalar curvature and the covariant derivative operator associated with $\gamma_{ij}$, respectively,\,\footnote{{The indices are raised and lowered---with the exception of expressions involving the covariant derivative $\mathbb{D}_i$---by $\hat\gamma^{ij}$ and $\hat\gamma_{ij}$, respectively. In the exceptional cases, involving $\mathbb{D}_i$, they will be raised and lowered by $\gamma^{ij}$ and $\gamma_{ij}$, respectively.}} whereas the two covariant derivative operators $\mathbb{D}_i$ and $\hat D_i$ are related by (4.11) and (4.12) of \cite{racz_geom_cauchy}.

\medskip

Note that as the $\Sigma_\sigma$ hypersurfaces in both cases are spacelike and, in virtue of (\ref{hij}), the metric $h_{ij}$ is Riemannian  in the secondary $1+[n-1]$ decomposition the correspondent of $\epsilon$ takes the value $+1$. In turn, most of the relations derived in subsection 4.2 of \cite{racz_geom_cauchy} remain intact. 
In particular, by applying the decomposition
\begin{equation}\label{dec_1_n}
K_{ij}= \boldsymbol\kappa \,\hat n_i \hat n_j  + \left[\hat n_i \,{\rm\bf k}{}_j  + \hat n_j\,{\rm\bf k}{}_i\right]  + {\rm\bf K}_{ij}\,,
\end{equation}
the relations 
\begin{eqnarray}
\hskip-.2cm\boldsymbol\kappa {}&\hskip-.2cm=&\hskip-.2cm{} \hat n^k\hat  n^l\,K_{kl} = \mycal{L}_n \ln \hat N \label{kappa_n}\\
\hskip-.2cm{\rm\bf k}{}_{i} {}&\hskip-.2cm=&\hskip-.2cm{} {\hat \gamma}^{k}{}_{i}\hat  n^l\, K_{kl} = {(2\hat N)}^{-1}\,\hat \gamma_{il}\,(\mycal{L}_{n}\hat N^l)  \label{loc_expr42_n}\\
\hskip-.2cm{\rm\bf K}_{ij} {}&\hskip-.2cm=&\hskip-.2cm{} {\hat \gamma}^{k}{}_{i} {\hat \gamma}^{l}{}_{j}\,K_{kl} = \tfrac12\,\hat \gamma^k{}_i\hat \gamma^l{}_j\,(\mycal{L}_{n}\hat \gamma_{kl})\, \label{loc_expr43_n}
\end{eqnarray}
and
\begin{equation}\label{trbfK}
{\rm\bf K}^l{}_{l}=\hat\gamma^{kl}\,{\rm\bf K}_{kl}={\tfrac12}\,\hat\gamma^{ij}(\mycal{L}_n\hat\gamma_{ij})=\tfrac{n-1}2\,\mathcal{L}_{n} \ln\Omega^2\,
\end{equation}
can be seen to hold. 
One should not forget, however, about that in the present case the surfaces $\mycal{S}_{\sigma,\rho}$ are $[n-1]$-dimensional thereby the trace of $\hat\gamma_{ij}$ is $n-1$ and some of the coefficients have to be adjust accordingly. For instance, the (conformal invariant) projection operator reads as
\begin{equation}\label{PI_def_n}
\Pi^{kl}{}_{ij}=\hat\gamma^{k}{}_{i}\hat\gamma^{l}{}_{j}-\tfrac1{n-1}\,\hat\gamma_{ij}\hat\gamma^{kl}=\gamma^{k}{}_{i}\gamma^{l}{}_{j}-\tfrac1{n-1}\,\gamma_{ij}\gamma^{kl}\,
\end{equation}
and the trace-free projection of ${\rm\bf K}_{ij}$ and the extrinsic curvature ${\hat K}_{ij}$ can be given as
\begin{equation}\label{interiors0_n}
\interior{\rm\bf K}_{ij}={\rm\bf K}_{ij}-\tfrac1{n-1}\,\gamma_{ij}(\gamma^{ef}{{\rm\bf K}}_{ef}) \ \ \ {\rm and }\ \ \ 
\interior{\hat K}{}_{ij}={\hat K}_{ij}-\tfrac1{n-1}\,\gamma_{ij}(\gamma^{ef}{{\hat K}}_{ef})\,.
\end{equation}

\section{The $1+n$ constraints}\label{constraints} 
\setcounter{equation}{0}

Analogously what was done in \cite{racz_geom_cauchy} the momentum constraint can be seen to be equivalent to   
\begin{align}
({\hat K^{l}}{}_{l})\,{\rm\bf k}{}_{i} + \hat D^l \interior{\rm\bf K}{}_{li} + \boldsymbol\kappa\,\dot{\hat n}{}_i +\mycal{L}_{\hat n} {\rm\bf k}{}_{i} - \dot{\hat n}{}^l\,{\rm\bf K}_{li} 
- \hat D_i\boldsymbol\kappa - \tfrac{n-2}{n-1}\,\hat D_i ({\rm\bf K}^l{}_{l}) -\epsilon\,\mathfrak{p}_l\,{\hat \gamma^{l}}{}_{i} = &{} 0 \label{par_const_n} \\
\boldsymbol\kappa\,({\hat K^{l}}{}_{l}) + \hat D^l {\rm\bf k}_{l} - {\rm\bf K}{}_{kl}{\hat K}{}^{kl}  -2\,\dot{\hat n}{}^l\, {\rm\bf k}_{l}  - \mycal{L}_{\hat n}({\rm\bf K}^l{}_{l})-\epsilon\,\mathfrak{p}_l\,{\hat n^{l}} = &{} 0\,, \label{ort_const_n}
\end{align}
{where $\dot{\hat n}{}_k={\hat n}{}^lD_l{\hat n}{}_k=-{\hat D}_k(\ln{\hat N})$.}

\medskip

By applying (5.3) and (5.4) of \cite{racz_geom_cauchy} we also get 
\begin{align}
\hskip-0.5cm({\hat K^{l}}{}_{l})\,{\rm\bf k}{}_{i} + \hat D^l \interior{\rm\bf K}{}_{li} + \boldsymbol\kappa\,\dot{\hat n}{}_i +{(2 \hat{N})}^{-1}{\hat{\gamma}_{ij}\,[\mathcal{L}_{\hat{n}}(\mathcal{L}_{n}\hat{N}^{j})]} + 2\,\hat{K}_{ij}\mathbf{k}^{j}- \mycal{L}_{\hat{n}}(\ln\hat{N}) \mathbf{k}_{i}   & \nonumber\\
- \dot{\hat n}{}^l\,{\rm\bf K}_{li} - \hat D_i\boldsymbol\kappa - \tfrac{n-2}2\,\hat D_i (\mathcal{L}_{n}\ln\Omega^2) -\epsilon\,\mathfrak{p}_l\,{\hat \gamma^{l}}{}_{i}{}&={} 0  \label{par_const2_n} \\
\boldsymbol{\kappa}\,\hat{K}^{l}{}_{l} - \mathbf{K}_{kl} \hat{K}^{kl} 
 + {(2 \hat{N})}^{-1}{\hat{D}_{l}(\mathcal{L}_{n}\hat{N}^{l})}
 - \tfrac{n-1}2 \mathcal{L}_{\hat{n}} [\mathcal{L}_{{n}}(\ln\Omega^2)]+ \mathbf{k}^{l}\,\hat{D}_{l}(\ln\hat{N}) -\epsilon\,\mathfrak{p}_{l}\hat{n}^{l} 
{}&={} 0.\label{ort_const2_n}
\end{align}

Notice that only the source terms---the last terms on the r.h.s.~of (\ref{par_const2_n}) and (\ref{ort_const2_n})---pick up an $\epsilon$ factor which means that the principal parts of the above equations are insensitive to the signature of the primary space. Remarkably, $\tfrac2{n-2}$ times of (\ref{par_const2_n}) and $-\tfrac{4\,\hat N}{n-1}$ times of (\ref{ort_const2_n})---when writing them out in coordinates $(\rho,x^3,\dots,x^{n+1})$ adopted to the foliation $\mycal{S}_{\sigma,\rho}$ and the vector field $\rho^i$ on $\Sigma_\sigma$---can be seen to take the form 
\begin{equation}\label{constr_hyp}
\left\{\hskip-.09cm \left(\hskip-.09cm
\begin{array}{cc}
 \hskip-.09cm \frac1{(n-2)\,\hat N{}^2}{\hat \gamma}_{AB} & \hskip-.19cm 0 \\ 
 \hskip-.19cm 0  &\hskip-.19cm 2 
\end{array} 
\hskip-.15cm \right)\hskip-.12cm\,\partial_\rho +
\left(\hskip-.09cm
\begin{array}{cc}
 \hskip-.19cm -  \frac{\hat N^{K}}{(n-2)\,\hat N{}^2}{\hat \gamma}_{AB} & \hskip-.19cm- {\hat \gamma}_A{}^{K}  \\
 \hskip-.19cm - {\hat \gamma}_B{}^{K} &  \hskip-.19cm -  2\,\hat N^{K}
\end{array} \hskip-.25cm
\right)\hskip-.12cm\,\partial_K\hskip-.09cm
\right\}
\left(\hskip-.19cm
\begin{array}{c}
\mathcal{L}_{n}\hat{N}^{B} \\
\mathcal{L}_{n} \ln\Omega^2
\end{array} \hskip-.19cm
\right) +
\left(\hskip-.19cm
\begin{array}{c}
{}^{{}^{(\mathcal{L}_{n}\hat{N}^{B})}}\hskip-1mm\mycal{B}_{\,A} \\
{}^{{}^{(\mathcal{L}_{n} \ln\Omega^2)}}\hskip-1mm\mycal{B}
\end{array} \hskip-.19cm
\right)=0\,,
\end{equation}
As in (5.7) of \cite{racz_geom_cauchy} the coefficient{ matrices} of $\partial_\rho$ and $\partial_K$ are symmetric {such that} the coefficient of $\partial_\rho$ is also positive definite. Thereby (\ref{constr_hyp}) possesses the form of a first order symmetric hyperbolic system 
\begin{equation}\label{constr_hyp2}
\mathcal A^{(\rho)} \,\partial_\rho {\bf u} + \mathcal A^{(K)} \,\partial_K {\bf u}+ \mathcal B = 0 
\end{equation}
(with $K=3,\dots,n+1$) with the vector valued variable 
\begin{equation}\label{vector-valued}
{\bf u}=(\mathcal{L}_{n}\hat{N}^{B}, \mathcal{L}_{n} \ln\Omega^2)^T\,.
\end{equation}

\medskip

Notice that all the coefficients and the source terms in (\ref{constr_hyp}) may be considered as smooth fields defined on the level surfaces $\mycal{S}_{\sigma,\rho}$ exclusively. 


\bigskip

In turning to the Hamiltonian constraint recall first that by (A.1) of \cite{racz_geom_det} with $\epsilon=+1$
\begin{equation}
{}^{{}^{(n)}}\hskip-1mm R= \hat R - \left\{2\,\mycal{L}_{\hat n} ({\hat K^l}{}_{l}) + ({\hat K^{l}}{}_{l})^2 + \hat K_{kl} \hat K^{kl} + 2\,{\hat N}^{-1}\,\hat D^l \hat D_l \hat N \right\}
\end{equation}
holds. By combining this with a suitable adaptation of (4.22)-(4.34) of \cite{racz_geom_cauchy} twice of the Hamiltonian constraint (\ref{expl_eh}) can be {put into} the form 
\begin{align}
-\epsilon\,\hat R + \epsilon\left\{2\,\mycal{L}_{\hat n} ({\hat K^l}{}_{l}) + ({\hat K^{l}}{}_{l})^2 + {\hat K}{}_{kl}\,{\hat K}{}^{kl} +  2\,{\hat N}^{-1}\,\hat D^l \hat D_l \hat N \right\} {}& + 2\,\boldsymbol\kappa\,({\rm\bf K}^l{}_{l})+({\rm\bf K}^l{}_{l})^2 \nonumber \\ {}& \hskip-2cm
-2\,{\rm\bf k}{}^{l}{\rm\bf k}{}_{l}  - {\rm\bf K}{}_{kl}\,{\rm\bf K}{}^{kl} -2\,\mathfrak{e}=0\,. \label{scal_constr_n}
\end{align}

Taking into account that the constraint equations are always underdetermined, along with the fact that by making use of (\ref{constr_hyp}) the `time' derivative of $\hat{N}^{A}$ and $\Omega$ [see (\ref{vector-valued}) above] can be determined, it would be advantageous if (\ref{scal_constr_n}) could be solved for the `time' derivative of $\hat{N}$. In virtue of (\ref{kappa_n}) this can {always} be done {if} (\ref{scal_constr_n}) can be solved for $\boldsymbol\kappa=\mycal{L}_{n} \ln\hat N$. Since $\boldsymbol\kappa$ is involved merely algebraically in (\ref{scal_constr_n}) whenever this can be done the Hamiltonian constraint is solved in a convenient way. In this respect it is of critical importance to know whether the coefficient ${\rm\bf K}^e{}_{e}=\tfrac{n-1}2\,\mathcal{L}_{n} \ln\Omega^2$ of $\boldsymbol\kappa$ in (\ref{scal_constr_n}) may vanish on open subsets of the base manifold. 

\medskip

In proceeding recall first that in the {four-dimensional} Lorentzian case there are configurations for which the vanishing of ${\rm\bf K}^e{}_{e}$ does indeed occur. For instance, a temporary vanishing happens at the turning point of a closed FLRW cosmological model. As a more annoying example one may also think of the cylindrical wave solution studied in \cite{ehlers, stachel, diverno1}) where ${\rm\bf K}^e{}_{e}$ is zero identically. Note, however, that these cylindrical wave solutions are non-generic and it is plausible to assume that this type of vanishing {of ${\rm\bf K}^e{}_{e}$} is excluded for generic Riemannian or Lorentzian spaces. As a partial support of this idea note first that in virtue of (\ref{trbfK})  ${\rm\bf K}^e{}_{e}$ measures the rate of change of the volume element---associated with the induced metric $\hat\gamma_{kl}$---on the $\mycal{S}_{\sigma,\rho}$ surfaces in the direction of the `time' evolution vector field $\sigma^a$. Whence the vanishing of ${\rm\bf K}^e{}_{
e}$, on open subsets of the base manifold, cannot occur unless 
the invariance of the volume element with respect to the flow determined by $n^a$ 
is guaranteed. 
As this type of invariance appears to be exceptional hereafter we shall use the following:

\begin{conjecture}\label{conj}
In a generic Riemannian or Lorentzian space, as specified in Sections \ref{prelim} and \ref{spec_choice_n}, the contraction ${\rm\bf K}^e{}_{e}$ does not vanish on open domains of $M$.  If it vanishes at certain isolated locations then (\ref{scal_constr_n}), combined with the l'Hospital rule, may still be used to determine the value of $\boldsymbol\kappa=\mycal{L}_{n} \ln\hat N$ there. 
\end{conjecture}

According to the discussion above whenever this conjecture holds the Hamiltonian constraint can be solved as an algebraic equation for $\mycal{L}_{n} \ln\hat N$. In the rest of this paper we shall assume that Conjecture \ref{conj} holds though no attempt will be made here to {verify} its validity.\,\footnote{It worth mentioning that in case of a generic setup---i.e.~whenever the primary shift vector is not required to vanish---the Hamiltonian constraint relevant for a time slicing of Schwarzschild spacetime in the Kerr-Schild form has nowhere vanishing trace ${\rm\bf K}^e{}_{e}$. This allows one to solve (\ref{scal_constr_n}) as an algebraic equation for $\boldsymbol\kappa$, e.g.~in a wide class of near Schwarzschild configurations.\label{schw}}


\bigskip

In proceeding note first that the solubility of the $1+n$ constraints is always of crucial significance. In this respect it is worth mentioning that (\ref{constr_hyp}) as a hyperbolic system can always be solved as an initial value problem with initial data specified at 
some $\mycal{
S}_\sigma\subset\Sigma_\sigma$ for the variables $ \mycal{L}_n\Omega, \mycal{L}_n \hat N^{A}$.
In solving this initial value problem the other dependent variables, i.e.~$\Omega, \gamma_{AB}$, $\hat N, \hat N^A, \mycal{L}_n\,\gamma_{AB}$ are  all  freely specifiable on $\Sigma_\sigma$ whereas $\mycal{L}_n\hat N$ is determined by (\ref{scal_constr_n}). 

\medskip

In summarizing we have:

\begin{theorem}\label{concl}
In terms of the geometrically distinguished variables the Hamiltonian and momentum constraints can be given as an algebraic-hyperbolic system comprised by (\ref{scal_constr_n}) and (\ref{constr_hyp}). {If} Conjecture \ref{conj} holds the algebraic-hyperbolic system can be solved on the hypersurfaces $\Sigma_\sigma$ for 
\begin{equation}
\mycal{L}_n\hat N, \mycal{L}_n\Omega, \mycal{L}_n \hat N^{A}
\end{equation}
{provided that} a sufficiently regular choice for the variables
\begin{equation}\label{list-free_a}
\Omega, \gamma_{AB}, \hat N, \hat N^A, \mycal{L}_n\,\gamma_{AB}
\end{equation}
had been made throughout $\Sigma_\sigma$. {The} initial data to the hyperbolic system (\ref{constr_hyp}) {can also be freely specified} on one of the {codimension-}two-surfaces $\mycal{S}_{\sigma,\rho}$ foliating $\Sigma_\sigma$.
\end{theorem}

\section{The evolutionary system}\label{evolutionary} 
\setcounter{equation}{0}

In proceeding consider first the system comprised by (\ref{ham_1+(1+2_n)})-(\ref{evol_1+(1+2_n)}).  The secondary `Hamiltonian and momentum' constraints, (\ref{ham_1+(1+2_n)}) and (\ref{mom_1+(1+2_n)}), read as 
\begin{eqnarray}  
\hat E^{{}^{(\mathcal{H})}} {}&\hskip-.4cm=&\hskip-.2cm{} \tfrac12\,\{ -\hat R + ({{\hat K}{}^{l}}{}_{l})^2 - \hat K_{kl}\hat K^{kl} - 2\,\hat{\mathfrak{e}} \}=0\,,\label{ham_1+(1+2)_detailed_n} \\
\hat E^{{}^{(\mathcal{M})}}_i {}&\hskip-.4cm=&\hskip-.2cm{} \hat D^l {{\hat K}}{}_{li} - \hat D_i {{\hat K}{}^{l}}{}_{l} - \hat{\mathfrak{p}}_{i}=0\,.\label{mom_1+(1+2)_detailed_n}
\end{eqnarray}

Turning to the secondary `evolution equation' $\hat E^{{}^{(\mathcal{EVOL})}}_{ij}=0$ note first that by making use of the (conformal invariant) projection operator (\ref{PI_def_n}) $\hat E^{{}^{(\mathcal{EVOL})}}_{ij}$ may be decomposed as
\begin{equation}\label{dec_evol}
\hat E^{{}^{(\mathcal{EVOL})}}_{ij}= \Pi^{kl}{}_{ij}\,\hat E^{{}^{(\mathcal{EVOL})}}_{kl} + \tfrac1{n-1}\,\hat\gamma_{ij}\,(\hat\gamma^{kl}\,\hat E^{{}^{(\mathcal{EVOL})}}_{kl}) \,,
\end{equation}
where---by repeating the analogous argument of Subsection 6.1 of \cite{racz_geom_cauchy}---the first term and the contraction in the second term can be seen to read as  
\begin{eqnarray}\label{evol_1+(1+2)_trf_1_n}
\Pi^{kl}{}_{ij}\,\hat E^{{}^{(\mathcal{EVOL})}}_{kl}=-\Pi^{kl}{}_{ij}\,\left[\mycal{L}_{\hat n}{\hat K}_{kl}+ ({\hat K}{}^m{}_{m})\,{\hat K}_{kl}+ \hat N{}^{-1}\hat D_k \hat D_l  \hat N + \hat{\mathfrak{S}}_{kl}\right]\,,
\end{eqnarray}
\begin{align}\label{evol_1+(1+2)_tr_1_n}
\hat \gamma^{kl}\,\hat E^{{}^{(\mathcal{EVOL})}}_{kl} = (n-2)\,\hat E^{{}^{(\mathcal{H})}} 
- \hat{\mathfrak{S}}_{kl}\,\hat \gamma^{kl}+\mycal{L}_{\hat n}({\hat K}{}^l{}_{l}) + \hat K_{kl}\hat K^{kl}+ \hat N{}^{-1}\hat D_k \hat D_l  \hat N \,, 
\end{align}
respectively. The source terms $\hat{\mathfrak{e}}, \hat{\mathfrak{p}}_{i}, \hat \gamma^{kl}\,\hat{\mathfrak{S}}_{kl}, \Pi^{kl}{}_{ij}\hat{\mathfrak{S}}_{kl}$ in (\ref{ham_1+(1+2)_detailed_n})-(\ref{evol_1+(1+2)_tr_1_n}) are defined as $\hat{\mathfrak{e}}=\hat n^k \hat n^l\,{}^{{}^{(n)}}\hskip-1mm \mycal{G}_{kl}$, $\hat{\mathfrak{p}}_{i}={{\hat \gamma}^{k}}{}_{i}\,\hat n^l\, {}^{{}^{(n)}}\hskip-1mm \mycal{G}_{kl}$, $\hat \gamma^{kl}\,\hat{\mathfrak{S}}_{kl}=\hat \gamma^{kl}\,[{{\hat \gamma}^{p}}{}_{k} {{\hat \gamma}^{q}}{}_{l}\,\,{}^{{}^{(n)}}\hskip-1mm \mycal{G}_{pq}]=\hat \gamma^{kl}\,{}^{{}^{(n)}}\hskip-1mm \mycal{G}_{kl}$ and $\Pi^{kl}{}_{ij}\hat{\mathfrak{S}}_{kl}=\Pi^{kl}{}_{ij}\,[{{\hat \gamma}^{p}}{}_{k} {{\hat \gamma}^{q}}{}_{l}\,{}^{{}^{(n)}}\hskip-1mm \mycal{G}_{pq}]=\Pi^{kl}{}_{ij}\,{}^{{}^{(n)}}\hskip-1mm \mycal{G}_{
kl}$, respectively. {As these source terms store the second order $\sigma$-derivatives of the basic variables they do not vanish, not even in case of the pure vacuum problem with $\mycal{G}_{ab}=0$.}

\bigskip

In attempting to solve the field equations it is always advantageous to explore their {latent} relations. In metric theories of gravity the Bianchi identity does always provide us the needed relations. In proceeding recall first that according to Lemma 3.2 of \cite{racz_geom_det}---it was also derived by making use of the Bianchi identity---whenever the $1+n$ constraint expressions $E^{{}^{(\mathcal{H})}}$ and $E^{{}^{(\mathcal{M})}}_a$ vanish on all the $\sigma=const$ level surfaces then the evolutionary expression $E^{{}^{(\mathcal{EVOL})}}_{ab}\hskip-.1cm{}$ must be subject to the relations 
\begin{eqnarray} 
K^{ab}\,E^{{}^{(\mathcal{EVOL})}}_{ab}\hskip-.1cm{}&\hskip-.2cm=&\hskip-.2cm{} 0\,, \label{cons_law1} \\ 
D^aE^{{}^{(\mathcal{EVOL})}}_{ab}\hskip-.1cm{} - \epsilon\,\dot n^a\,E^{{}^{(\mathcal{EVOL})}}_{ab}\hskip-.1cm{}  {}&\hskip-.2cm=&\hskip-.2cm{} 0\,, \label{cons_law2}
\end{eqnarray}
where $\dot n_a:=n^e\nabla_e n_a=-\epsilon\,D_a \ln N$.

\medskip

Note also that, analogously to (2.5) of \cite{racz_geom_det}, $E^{{}^{(\mathcal{EVOL})}}_{ab}$ may be decomposed as
\begin{align}\label{axu}
{{}^{{}^{(n)}}\hskip-1mm E_{ab}} \hskip-.1cm{}=\hat E^{{}^{(\mathcal{H})}}\hat n_a\hat n_b + [ \hat n_a\hat E^{{}^{(\mathcal{M})}}_b\hskip-.1cm{} +\hat n_b\hat E^{{}^{(\mathcal{M})}}_a ] + (\hat E^{{}^{(\mathcal{EVOL})}}_{ab}\hskip-.1cm{} + \hat\gamma_{ab}\hat E^{{}^{(\mathcal{H})}} )\,.
\end{align}
{Note also that ${}^{{}^{(n)}}\hskip-1mm E_{ab}$ can be seen to be equal---modulo the Hamiltonian constraint times $h_{ab}$---to $E^{{}^{(\mathcal{EVOL})}}_{ab}$. As the primary Hamiltonian constraint is supposed to vanish in the present case, (\ref{cons_law1}), (\ref{cons_law2}) and (\ref{axu})}, along with (\ref{dec_1_n}), imply that
\begin{align}
K^{ab}E^{{}^{(\mathcal{EVOL})}}_{ab}\hskip-.1cm{} {}&=\boldsymbol\kappa\,\hat E^{{}^{(\mathcal{H})}}  + 2\,{\rm\bf k}{}^{e}\hat E^{{}^{(\mathcal{M})}}_e + {\rm\bf K}{}^{ef}\,\hat E^{{}^{(\mathcal{EVOL})}}_{ef}\hskip-.1cm{} + ({\rm\bf K}{}^{e}{}_e)\,\hat E^{{}^{(\mathcal{H})}} \label{cons_law11}\\  
\dot n^a E^{{}^{(\mathcal{EVOL})}}_{ab}\hskip-.1cm{} {}&=[(\hat n_a \dot n^a)\hat E^{{}^{(\mathcal{H})}} + (\dot n^a\,\hat E^{{}^{(\mathcal{M})}}_a )]\hat n_b + (\hat n_a \dot n^a) \hat E^{{}^{(\mathcal{M})}}_b + \dot n^a\,[\hat E^{{}^{(\mathcal{EVOL})}}_{ab}\hskip-.1cm{} + \hat\gamma_{ab}\,  \hat E^{{}^{(\mathcal{H})}}]\,.\label{cons_law22}
\end{align}
Note also that by adopting (A.9) and (A.10) of \cite{racz_geom_det} the parallel and orthogonal projections of $D^{a}E^{{}^{(\mathcal{EVOL})}}_{ab}$ may be put into the form 
\begin{align}
{} \hat n^e{}D^{a}E^{{}^{(\mathcal{EVOL})}}_{ae}\hskip-.1cm{}  {} &=\mycal{L}_{\hat n}\,\hat E^{{}^{(\mathcal{H})}}\hskip-.1cm{}+ \hat D^e \hat E^{{}^{(\mathcal{M})}}_e + (\hat K^e{}_e) \,\hat E^{{}^{(\mathcal{H})}}\hskip-.1cm{} - [\hat E^{{}^{(\mathcal{EVOL})}}_{eb}\hskip-.1cm{}+\hat\gamma_{eb}\,\hat E^{{}^{(\mathcal{H})}}]\,\hat K^{ef}   - 2\,\dot {\hat n}^e  \hat E^{{}^{(\mathcal{M})}}_e  \\
{} \hat\gamma^e{}_b D^{a}E^{{}^{(\mathcal{EVOL})}}_{ae}\hskip-.1cm{} {}&=\mycal{L}_{\hat n}\,\hat E^{{}^{(\mathcal{M})}}_b+\hat D^e [\hat E^{{}^{(\mathcal{EVOL})}}_{eb}\hskip-.1cm{}+\hat\gamma_{eb}\,\hat E^{{}^{(\mathcal{H})}}]+(\hat K^e{}_e) \,\hat E^{{}^{(\mathcal{M})}}_b - \dot {\hat n}^e\, \hat E^{{}^{(\mathcal{EVOL})}}_{eb} \,.
\end{align} 

\medskip

Assume now that, in addition to the $1+n$ constraints, the (conformal invariant) projection of the secondary evolution equation holds, i.e.
\begin{equation}\label{dec_evol_2}
\Pi^{kl}{}_{ij}\,\hat E^{{}^{(\mathcal{EVOL})}}_{kl} =0\,.
\end{equation}
Then, in virtue of (\ref{dec_evol}), (\ref{cons_law1}), (\ref{cons_law11})  and (\ref{dec_evol_2})
\begin{align}
K^{ab}E^{{}^{(\mathcal{EVOL})}}_{ab}\hskip-.1cm{} {}&=(\boldsymbol\kappa+{\rm\bf K}{}^{e}{}_e)\,\hat E^{{}^{(\mathcal{H})}}  + 2\,{\rm\bf k}{}^{e}\hat E^{{}^{(\mathcal{M})}}_e + \tfrac1{n-1}\,({\rm\bf K}{}^{e}{}_e)\,(\hat \gamma^{kl}\,\hat E^{{}^{(\mathcal{EVOL})}}_{kl}) =0\,. \label{cons_law111}
\end{align}
It immediately follows that whenever Conjecture \ref{conj} holds this equation can be solved for the contraction $\hat \gamma^{kl}\,\hat E^{{}^{(\mathcal{EVOL})}}_{kl}$ such that the solution is linear and homogeneous in the expressions $\hat E^{{}^{(\mathcal{H})}}$ and $\hat E^{{}^{(\mathcal{M})}}_b$. By substituting the corresponding expression into (\ref{cons_law22}) and taking into account the parallel and orthogonal projections of (\ref{cons_law2}) a system of the form
\begin{align}
\mycal{L}_{\hat n}\,\hat E^{{}^{(\mathcal{H})}}\hskip-.1cm{}+ \hat\gamma^{ef} \hat D_e \hat E^{{}^{(\mathcal{M})}}_f = {}& \hat{\mycal{E}} \label{hyp_1}\\
\mycal{L}_{\hat n}\,\hat E^{{}^{(\mathcal{M})}}_b - ({{\rm\bf K}{}^{e}{}_e})^{-1}\,[\,\boldsymbol\kappa\,\hat D_b \hat E^{{}^{(\mathcal{H})}}  + 2\,{\rm\bf k}{}^{e}\hat D_b\hat E^{{}^{(\mathcal{M})}}_e] = {}&  \hat{\mycal{E}}_b\, \label{hyp_2}
\end{align}
can be seen to hold, where, in virtue of (\ref{cons_law11})-(\ref{cons_law111}), $\hat{\mycal{E}}$ and $\hat{\mycal{E}}_b$ are linear and homogeneous expressions in $\hat E^{{}^{(\mathcal{H})}}$ and $\hat E^{{}^{(\mathcal{M})}}_i$, respectively. 

\medskip

It can be verified by a direct calculation that these equations comprise a Friedrichs symmetrizable first order linear and homogeneous system of partial differential equations for the vector variable $(\hat E^{{}^{(\mathcal{H})}},\hat E^{{}^{(\mathcal{M})}}_i)^T$ provided that $\boldsymbol\kappa$ and ${{\rm\bf K}{}^{e}{}_e}$ are of opposite sign. It is important to be mentioned that this latter requirement is harmless as it holds, e.g.~for a time slicing  of Schwarzschild spacetime within the Kerr-Schild setup, mentioned in footnote \ref{schw}, therefore it is expected to hold for a wide class of near Schwarzschild configurations, as well. These type of  systems are known to have unique solutions (see e.g.~\cite{benzoni}) whence they possess the identically zero solution for vanishing initial data.
When this happens, in virtue of (\ref{cons_law111}), the vanishing of the contraction $\hat \gamma^{kl}\,\hat E^{{}^{(\mathcal{EVOL})}}_{kl}$ is also guaranteed on the $\Sigma_\sigma$ level surfaces. 

Note that whenever one intends to integrate (\ref{hyp_1}) and (\ref{hyp_2}) starting at an origin the irregular behavior of ${{\rm\bf K}{}^{e}{}_e}$ there, i.e.~the limiting behavior $\lim_{\rho\rightarrow \rho^{\pm}_*}\,{{\rm\bf K}{}^{e}{}_e}=\pm\infty$ guarantees to have suitable type of characteristics at $\rho=\rho_*$. 
It is also important to keep in mind that in the above argument concerning the solubility of (\ref{hyp_1}) and (\ref{hyp_2}) we tacitly assumed more than guaranteed by Conjecture \ref{conj}. Namely, the above argument presumes that ${{\rm\bf K}{}^{e}{}_e}$ nowhere vanishes. It would be important to know if the vanishing of ${{\rm\bf K}{}^{e}{}_e}$ at isolated locations could be compatible with the uniqueness of solutions to (\ref{hyp_1}) and (\ref{hyp_2}) required by the argument above.  

\medskip

Note that by combining what has been verified above with the existence of a regular origin at the line {$\mycal{W}_{\rho_*}$}, by an argument analogous to the one applied in \cite{racz_geom_cauchy}, it can be verified that to get solutions to the full set of Einstein equations it suffices to solve merely the reduced system comprised by (\ref{constr_hyp}), (\ref{scal_constr_n}) and (\ref{dec_evol_2}).   

To see that this is indeed the case note that by substituting the relations listed in (\ref{origin_n}) into (\ref{ham_1+(1+2)_detailed_n}) and (\ref{mom_1+(1+2)_detailed_n}), and taking the limit $\rho\rightarrow\rho_*$ both of the secondary constraint expressions $\hat E^{{}^{(\mathcal{H})}}$  and $\hat E^{{}^{(\mathcal{M})}}_i$ can be seen to vanish identically along the line {$\mycal{W}_{\rho_*}$} {representing a regular} origin {in $M$}. This implies then that the initial data for the first order symmetric hyperbolic system comprised by (\ref{hyp_1}) and (\ref{hyp_2}) vanish so $\hat E^{{}^{(\mathcal{H})}}, \hat E^{{}^{(\mathcal{M})}}_i$, along with $\hat \gamma^{kl}\,\hat E^{{}^{(\mathcal{EVOL})}}_{kl}$, vanish identically on each of the $\Sigma_\sigma$ level surfaces at $\rho=\rho_*$. Whence the remaining field equations to be solved are (\ref{constr_hyp}), (\ref{scal_constr_n}) and (\ref{dec_evol_2}). 

\medskip

In summarizing we have:  
\begin{theorem}\label{hypTh3}
Assume that Conjecture \ref{conj} holds and a regular origin exists in $M$. Then the secondary constraint expressions $\hat E^{{}^{(\mathcal{H})}}$  and $\hat E^{{}^{(\mathcal{M})}}_i$, along with the contraction $\hat \gamma^{kl}\,\hat E^{{}^{(\mathcal{EVOL})}}_{kl}$, vanish identically throughout $M$. Thus, regardless whether the primary space is Riemannian or Lorentzian, solutions to the reduced system comprised by (\ref{constr_hyp}), (\ref{scal_constr_n}) and (\ref{dec_evol_2}) are also solutions to the full set of Einstein's equations. 
\end{theorem}

\section{On the solubility of the reduced system}\label{reduced}
\setcounter{equation}{0}

In virtue of Theorem \ref{hypTh3} the solubility of the reduced system comprised by (\ref{constr_hyp}), (\ref{scal_constr_n}) and (\ref{dec_evol_2}) is of critical importance. 

\medskip

As it has already been discussed in Section \ref{constraints} (see Theorem \ref{concl} there) {if} suitable fields $\Omega, \gamma_{AB}, \hat N, \hat N^A, \mycal{L}_n\,\gamma_{AB}$ are chosen on the $\Sigma_\sigma$ level surfaces the algebraic-hyperbolic subsystem, comprised by  (\ref{scal_constr_n}) and (\ref{constr_hyp}), can be solved---as a well-posed initial value problem---for $\mycal{L}_n\hat N, \mycal{L}_n\Omega$ and $\mycal{L}_n \hat N^{A}$ on each of these surfaces.  

\medskip

From practical point of view once {suitable} fields $\Omega, \gamma_{AB}, \hat N, \hat N^A, \mycal{L}_n\,\gamma_{AB}$ are specified on the initial data surface $\Sigma_0$, by solving the constraints equations, the `time' derivatives $\mycal{L}_n\hat N, \mycal{L}_n\Omega$ and $\mycal{L}_n \hat N^{A}$  {can also be determined} there. {O}n the succeeding $\Sigma_\sigma$ level surfaces, the fields $\Omega, \hat N, \hat N^A$ can {then} be determined simply by integrating along the $\sigma^a=(\partial/\partial \sigma)^a$ `time lines'. In order to close this inductive process we also need to evaluate the fields $\gamma_{AB}$ and $\partial_\sigma\gamma_{AB}$ on the succeeding $\Sigma_\sigma$ level surfaces. {(\ref{dec_evol_2}) can be used to do so.} By applying a calculation analogous to the one used to determine the principal parts of the field 
equations (6.25)-(6.29) in \cite{racz_geom_cauchy}, the principal part of (\ref{dec_evol_2}) can be seen to take the form 
\begin{align}\label{princ_PI}
\tfrac12\,\gamma^k{}_i\gamma^l{}_j(\epsilon\, {}& \mycal{L}^2_{n}\gamma_{kl} +\mycal{L}^2_{\hat n}\gamma_{kl})  +\tfrac1{n-1}\,{\hat N}^{-1}\gamma_{ij}\,\mycal{L}_{\hat n}(\mathbb{D}_l \hat N^l) \\  {}& + \Omega^{-2}\,\Pi^{kl}{}_{ij}[ \mathbb{D}_k \mathbb{D}_l(\ln N +\ln \hat N) ] + \{\rm lower\ order\ terms \} =0 \,.\nonumber 
\end{align}
The simplicity of (\ref{princ_PI}) is remarkable. Besides, once the data $\Omega, \gamma_{AB}, \hat N, \hat N^A, \mycal{L}_n\,\gamma_{AB}$ for the constraints are known, the second and third terms in (\ref{princ_PI})---they contain second order derivatives of $\hat N$ and $\hat N^A$---can always be evaluated on each of the $\Sigma_\sigma$ level surfaces as the involved second order derivatives are all tangential to $\Sigma_\sigma$. Therefore, these terms do play, along with the lower order terms, the role of source terms which verifies that (\ref{princ_PI}) is indeed a second order partial differential equation for the conformal structure $\gamma_{ij}$.

\medskip

Remarkably when writing (\ref{princ_PI}) out in coordinates $(\sigma,\rho,x^3,\dots,x^{n+1})$ adopted to the foliation $\mycal{S}_{\sigma,\rho}$ and the vector fields $\sigma^i$ and $\rho^i$ it can be seen to take the form 
\begin{equation}\label{princ_PI_coord}
\left[ \frac{\epsilon}{\nu^2}\partial^2_\sigma+\partial^2_\rho -2\,{\hat N}^E\, \partial_\rho \partial_E + {\hat N}^E {\hat N}^F \partial_E \partial_F\right]\,{\gamma}_{AB} + {\mathcal F}_{AB}=0\,,
\end{equation}
where ${\mathcal F}_{AB}$ is a smooth functional of the variables $\Omega, \gamma_{AB}, \hat N, \hat N^A, \partial_\sigma\gamma_{AB}$,  their first order $\partial_\rho$ and $\partial_F$ derivatives and the second order $\partial_\rho\partial_F$ and $\partial_E \partial_F$ type derivatives of $\hat N$ and  $\hat N^A$. 

\medskip

As $\nu$ does not vanish and is finite, for the Lorentzian case, with $\epsilon=-1$,  (\ref{princ_PI}) possesses the form of a nonlinear wave equation which, by applying a standard order reduction procedure and the primary Hamiltonian constraint, can be put into a strongly hyperbolic first order system\,\footnote{{For the definition and properties of strongly hyperbolic system see, e.g.~\cite{reula}.}} {which} possesses a well-posed initial value problem for the conformal structure ${\gamma}_{AB}$.

\medskip

As already indicated (see footnote \ref{foot}) the gauge fixing {we apply} provides us an unexpected bonus concerning the solubility of  (\ref{geom_gd}) in case of spaces with Euclidean signature. To make this to be transparent start by replacing the formal `time coordinate' $\sigma$ by $\varsigma=i\,\sigma$, i.e. by applying the transformation 
\begin{equation}\label{trafo}
\sigma \to \varsigma=i\,\sigma\,.
\end{equation}
In the Riemannian case $\epsilon=1$ and the only elliptic equation to be solved for the conformal structure is (\ref{princ_PI}) which becomes formally hyperbolic under the action of (\ref{trafo}). We shall refer to the analogously {derived} equations by putting a box around their primary equation number. Accordingly, the equation yielded from (\ref{princ_PI}) will be denoted by \setlength\fboxrule{1pt}\setlength\fboxsep{0.1mm}\fbox{(\ref{princ_PI})}.

\medskip

Note that this formal hyperbolicity could become useless if the $1+n$ constraints would loose their preferable character, namely if their transformed form would not form a hyperbolic-algebraic system {any longer}. Remarkably the use of the proposed gauge fixing---see condition (1) in Section \ref{spec_choice_n}---guarantees that the transformed form of the $1+n$ constraints {retain their} distinguished features. To see this note first that the momentum constraint, given by (\ref{constr_hyp}), is linear and homogeneous in either the `time' derivative $\mycal{L}_n$ or in the normal vector `$n^a$'. The inspection of the terms involved in (\ref{par_const2_n}), (\ref{ort_const2_n}), and use the relations (\ref{kappa_n})-(\ref{loc_expr43_n}) verifies these claims. Analogously, the Hamiltonian constraint can be seen to be quadratic in these critical terms so only the sign of certain terms in the algebraic equation will be changed under the action of (\ref{trafo}). 
Consequently, the hyperbolic-algebraic character of the primary 
constraints will be retained by the new system comprised by \setlength\fboxrule{1pt}\setlength\fboxsep{0.1mm}\fbox{(\ref{constr_hyp})} and \setlength\fboxrule{1pt}\setlength\fboxsep{0.1mm}\fbox{(\ref{scal_constr_n})}. 

\medskip

To provide a slightly different wording of the above outlined argument recall first that our original Riemannian space can be represented by a pair of real fields $(h_{kl},K_{kl})$ such that various projections of $h_{kl}$ and $K_{kl}$ are subject to (5.5), (5.9) and (7.1). By applying (\ref{trafo})---in virtue of the vanishing of the primary shift vector---$K_{kl}=\tfrac12\,\mycal{L}_n h_{kl}$ {gets to be} replaced by the complex field $i K_{kl}$. By inserting the corresponding projections of $h_{kl}$ and $i K_{kl}$ into (\ref{constr_hyp}) the yielded equation \setlength\fboxrule{1pt}\setlength\fboxsep{0.1mm}\fbox{(\ref{constr_hyp})} {simply gets to be $i$ times of} the original first order symmetric hyperbolic system (\ref{constr_hyp}) since the latter is linear and homogeneous either in the `time' derivative $\mycal{L}_n$ or in the normal vector `$n^a$'. As the Hamiltonian constraint (\ref{scal_constr_n}) is 
quadratic in these critical terms only the sign 
of certain terms in 
\setlength\fboxrule{1pt}\setlength\fboxsep{0.1mm}\fbox{(\ref{scal_constr_n})} will differ from those in (\ref{scal_constr_n}). As equation (\ref{princ_PI}) is also quadratic in the critical terms it transforms the same way as the Hamiltonian constraint does whereas the principal part of \setlength\fboxrule{1pt}\setlength\fboxsep{0.1mm}\fbox{(\ref{princ_PI})} gets to be manifestly hyperbolic.

\medskip

It follows then that the pair of real fields $(h_{ij},K_{ij})$---representing the original (real) Riemannian space---are also solutions to the hyperbolic-algebraic-hyperbolic system comprised by  \setlength\fboxrule{1pt}\setlength\fboxsep{0.1mm}\fbox{(\ref{constr_hyp})}, \setlength\fboxrule{1pt}\setlength\fboxsep{0.1mm}\fbox{(\ref{scal_constr_n})}  and \setlength\fboxrule{1pt}\setlength\fboxsep{0.1mm}\fbox{(\ref{princ_PI})}. Note that the above process can also be reversed, i.e~it can be shown that any solution $(h_{ij},K_{ij})$ to the hyperbolic-algebraic-hyperbolic system comprised by \setlength\fboxrule{1pt}\setlength\fboxsep{0.1mm}\fbox{(\ref{constr_hyp})}, \setlength\fboxrule{1pt}\setlength\fboxsep{0.1mm}\fbox{(\ref{scal_constr_n})}  and \setlength\fboxrule{1pt}\setlength\fboxsep{0.1mm}\fbox{(\ref{princ_PI})} is also solution to the original system (\ref{constr_hyp}), (\ref{scal_constr_n}) and (\ref{princ_PI}).

\medskip

These observations imply then that to find solutions to (\ref{geom_gd}) in case of spaces with Euclidean signature one may always perform the transformation (\ref{trafo}) and aim to find solutions to the corresponding (non-physical) reduced hyperbolic-algebraic-hyperbolic system comprised by  \setlength\fboxrule{1pt}\setlength\fboxsep{0.1mm}\fbox{(\ref{constr_hyp})}, \setlength\fboxrule{1pt}\setlength\fboxsep{0.1mm}\fbox{(\ref{scal_constr_n})}  and \setlength\fboxrule{1pt}\setlength\fboxsep{0.1mm}\fbox{(\ref{princ_PI})}. 

\medskip

What we have proved in this section together with Theorem \ref{hypTh3} justifies

\begin{theorem}\label{hypThRiem}
Assume that Conjecture \ref{conj} holds and a regular origin exists in $M$. Then any solution to the reduced hyperbolic-algebraic-hyperbolic system comprised 
\begin{itemize}
\item  by (\ref{constr_hyp}), (\ref{scal_constr_n}) and (\ref{princ_PI}) in the Lorentzian signature case, or  
\item  by \setlength\fboxrule{1pt}\setlength\fboxsep{0.1mm}\fbox{(\ref{constr_hyp})}, \setlength\fboxrule{1pt}\setlength\fboxsep{0.1mm}\fbox{(\ref{scal_constr_n})} and \setlength\fboxrule{1pt}\setlength\fboxsep{0.1mm}\fbox{(\ref{princ_PI})} in the Euclidean signature case, 
\end{itemize}
is also solution to the full set of Einstein's equations in the respective cases.
\end{theorem}
 
\section{Final remarks}\label{final}
\setcounter{equation}{0}

In this paper $[n+1]$-dimensional ($n\geq 3$) smooth Einsteinian spaces of Euclidean and Lorentzian signature were investigated. Our principal concern was to determine the true degrees of freedom by restricting attention to spaces with base manifolds which can be foliated by a two-parameter family of homologous codimension-two-surfaces. This topological assumption allowed us to perform a pair of nested $1+n$ and $1+[n-1]$ decompositions.  By applying this, along with the notion of the conformal structure and the canonical form of the metric, a gauge fixing was defined. Remarkably for generic spaces fitting to the requirements imposed in Conjecture \ref{conj} and possessing a regular origin the followings could be verified: 
\begin{itemize}
\item The $1+n$ Hamiltonian and momentum constraints were found to form a coupled algebraic-hyperbolic system. It was shown that the hyperbolic part of this system forms a first order symmetric hyperbolic system which can be solved as a well-posed initial value problem. 
\item Concerning the full set of Einstein's equations it was shown that by combining the $1+n$ constraints with the trace-free projection of reduced evolutionary system of the secondary $1+[n-1]$ decomposition a reduced system can be formed. It was shown that solutions to this reduced system are also solutions to the full set of Einstein's equations in spaces possessing a regular origin. 
\item This reduced system was shown to form---in the Riemannian case after introducing an imaginary `time'---a reduced hyperbolic-algebraic-hyperbolic system. In the Riemannian case this {offers the possibility of developing a completely} new method for solving Einstein's equations.
\end{itemize}

Remarkably all of the above mentioned results apply---with negligible technical differences---to Einsteinian spaces {regardless of their} signature. In particular, if globally regular real solutions to Einstein's equation exists in the Riemannian case they have to be among those which, for suitable initial data choices, can be recovered by the hyperbolic method proposed in this paper.

\medskip

Concerning the geometrical degrees of freedom of Einstein's theory some preliminary insight had already been acquired in \cite{racz_geom_cauchy} by inspecting the freely specifiable variables on the initial data surface $\Sigma_0$. In this paper a dynamical identification of these degrees of freedom was attempt to be done by singling out a distinguished reduced system that was found to be equivalent to the full set of Einstein's equations for spaces with a regular origin and satisfying Conjecture \ref{conj}. The only truly evolutionary type of equation among the ones comprising this reduced system is the nonlinear wave equation (\ref{princ_PI_coord})---in the Riemannian case it becomes a wave equation after performing the transformation (\ref{trafo})---governing the dynamics of the conformal structure $\gamma_{AB}$. The initial data for this equation consists of $\gamma_{AB}$ and $\partial_\sigma\,\gamma_{AB}$ which is also part of the freely specifiable data (see (\ref{list-free_a}) in Theorem \ref{concl}) 
on 
$\Sigma_0$ for the $1+n$ constraints. It is also remarkable that the conformal structure is the only part of this freely specifiable data on $\Sigma_0$ that involves both a field and its `time' derivative.  All in all, these two aspects are more then suggestive in concluding that $\gamma_{AB}$ does indeed provide a convenient embodiment of the true gravitational degrees of freedom of the considered theories.    

\medskip

It may also be of interest to {determine the number of the} degrees of freedom $f$ {(per point of $M$) as a function} of dimension $dim(M)=n+1$. To this end recall first that the induced metric $\hat\gamma_{kl}$ has $\frac{(n-1)\,n}{2}$ independent components.  As the conformal structure $\gamma_{kl}$ is yielded from  $\hat\gamma_{kl}$ by imposing the algebraic restriction (\ref{eta}) we have that $f=\frac{(n-1)\,n}{2}-1$ as tabulated in Table \ref{Table1}. 
\begin{table}
\begin{center}
\begin{tabular}{|c||c|c|c|c|c|}
\hline
$dim(M)$ & 4 & 5 & 6 & $\dots$ & n+1 \\
\hline
$f$ & 2 & 5 & 9 & $\dots$ & $\frac{(n-1)\,n}{2}-1$ \\
\hline
\end{tabular}
\end{center}
\caption{\footnotesize The number of the gravitational degrees of freedom $f$ {(per point of $M$)} as a function of {dimension} $dim(M)$.}
\label{Table1}
\end{table}

\medskip

One should also keep in mind the limitations of the presented new results. Analogously to what we had in \cite{racz_geom_det,racz_geom_cauchy}, in deriving them considerations were restricted exclusively to the geometric part {by applying} Einstein's equations in the form (\ref{geom_gd}). Due to the {presence} of the generic source term $\mycal{G}_{ab}$ in (\ref{geom_gd}), our conclusions concerning the gravitational degrees of freedom will remain intact. Nevertheless,  whenever matter fields are included{---in studying the coupled evolutionary system---}careful case by case investigations have to {be applied.} 

\medskip
The specific form of the evolution equation (\ref{princ_PI}) governing the evolution of the conformal structure {is more than suggestive} that the conformal structure, as a fundamental variable, provides a natural basis for a reduced configuration space for the considered Einsteinian spaces. One of the most important related open issues is whether this {embodiment} of the degrees of freedom of gravity can be utilized in quantizing general relativity. As noted in \cite{ashtekar&geroch}, to have a natural set up for canonical quantization, i.e.~constructing a phase which is a cotangent bundle over the unconstrained configuration space, the identification of the true dynamical degrees of freedom is of critical significance. In this respect it would be of great importance to know whether the desired cotangent bundle can be developed by starting with the conformal structure. 

\section*{Acknowledgments}

The author is grateful to Lars Anderson, Ingemar Bengtsson and Bob Wald for helpful comments and suggestions. 
This research was supported by the European Union and the State of Hungary, co-financed by the European Social Fund in the framework of T\'AMOP-4.2.4.A/2-11/1-2012-0001 ``National Excellence Program''. 


\end{document}